\documentclass[12pt]{iopart}
\usepackage{color}
\usepackage{graphics}
\usepackage{pstricks}
\usepackage{ifthen}
\newpsobject{showgrid}{psgrid}{subgriddiv=1}     
\newcommand{\EventSet}[1]{ \ifthenelse{ #1 = 1}{\{N\}}{%
                           \ifthenelse{ #1 = 2}{\{E\}}{%
                           \ifthenelse{ #1 = 3}{\{N,E\}}{%
                           \ifthenelse{ #1 = 4}{\{S\}}{%
                           \ifthenelse{ #1 = 5}{\{N,S\}}{%
                           \ifthenelse{ #1 = 6}{\{E,S\}}{%
                           \ifthenelse{ #1 = 7}{\{N,E,S\}}{%
                           \ifthenelse{ #1 = 8}{\{W\}}{%
                           \ifthenelse{ #1 = 9}{\{N,W\}}{%
                           \ifthenelse{ #1 = 10}{\{E,W\}}{%
                           \ifthenelse{ #1 = 11}{\{N,E,W\}}{%
                           \ifthenelse{ #1 = 12}{\{S,W\}}{%
                           \ifthenelse{ #1 = 13}{\{N,S,W\}}{%
                           \ifthenelse{ #1 = 14}{\{E,S,W\}}{%
                           \ifthenelse{ #1 = 15}{\{N,E,S,W\}}{%
                           \ifthenelse{ #1 = -1}{}{%
                            *** undefined set type #1*** %
                           }}}}}}}}}}}}}}}}}
\newcommand{\EventYX}[2]{ \EventTypeY{#1}_{#2} }
\newcommand{\EventY}[2]{\EventYX{#1}{\EventSet{#2}}}
\newcommand{\EventS}[1]{\EventY{S}{#1}}
\newcommand{\EventU}[1]{\EventY{U}{#1}}
\newcommand{\EventTypeY}[1]{{\mathsf #1}}
\newcommand{\EventType}{\EventTypeY{T}}
\newcommand{\Spanning}{\EventTypeY{I}}
\newcommand{\Wrapping}{\EventTypeY{O}}
\newcommand{\Crossing}{\EventTypeY{X}}
\newcommand{\SpWrAdd}{+}
\newcommand{\SpanWrap}{\Spanning^\SpWrAdd}
\newcommand{\DblCross}{\Crossing^\SpWrAdd}
\newcommand{\WrapSpan}{\Wrapping^\SpWrAdd}
\newcommand{\exact}[1]{\widehat{#1}}
\newcommand{\histo}{{\mathcal P}}
\newcommand{\histobar}{{\exact{\mathcal P}}}
\newcommand{\Moment}{{\mathcal M}}
\newcommand{\Momentbar}{{\exact{\Moment}}}
\newcommand{\Ssite}{{\textrm{\tiny(s)}}}
\newcommand{\Sbond}{{\textrm{\tiny(b)}}}
\newcommand{\psite}{p^\Ssite}
\newcommand{\pbond}{p^\Sbond}
\newcommand{\AmpCrossing}{C_\Crossing}
\newcommand{\AmpDblCross}{C_\DblCross}
\newcommand{\AmpSpanning}{C_\Spanning}
\newcommand{\AmpSpanWrap}{C_\SpanWrap}
\newcommand{\AmpWrapping}{C_\Wrapping}
\newcommand{\AmpWrapSpan}{C_\WrapSpan}
\newcommand{\half}{\frac{1}{2}}
\newcommand{\quarter}{\frac{1}{4}}
\newcommand{\OC}{{\mathcal O}}
\begin{document}

\title[Crossing, spanning and wrapping in two-dimensional percolation]
{Numerical results for crossing, spanning and wrapping in 
two-dimensional percolation}

\author{Gunnar Pruessner$\dagger$ and Nicholas R Moloney$\ddagger$}

\address{$\dagger$Department of Mathematics,
Imperial College London,
180 Queen's Gate,
London SW7~2BZ,
UK}

\address{$\ddagger$Beit Fellow for Scientific Research,
Condensed Matter Theory,
Blackett Laboratory,
Imperial College London,
Prince Consort Rd,
London SW7~2BW,
UK}
\date{\today}

\ead{gunnar.pruessner@physics.org}

\begin{abstract}
Using a recently developed method to simulate percolation on large
clusters of distributed machines \cite{MoloneyPruessner:2003},
we have numerically calculated crossing, spanning and wrapping
probabilities in two-dimensional site and bond percolation with
exceptional accuracy.  Our results are fully consistent with predictions
from Conformal Field Theory.  We present many new results that await
theoretical explanation, particularly for wrapping clusters on a cylinder.  
We therefore provide possibly the most up-to-date reference for
theoreticians working on crossing, spanning and wrapping probabilities
in two-dimensional percolation.
\end{abstract}

\submitto{\JPA}
\pacs{64.60.Ak, 05.70.Jk}
\maketitle

\section{Introduction}
In the last decade, percolation has enjoyed the attention of conformal
field theorists who have sought to calculate crossing probabilities for
various aspect ratios and geometries. In rough terms, calculations
generally involve mapping percolation to a 1-state Potts model,
constructing a correlation function corresponding to the boundary
conditions necessitated by the crossing cluster of interest, and finding
a differential equation thereof. In 1991, Langlands \etal
\cite{LanglandsPichetPouliotStAubin:1992} were the first to investigate
``the universality of crossing probabilities in two-dimensional
percolation''. Shortly afterwards, Cardy \cite{Cardy:1992} obtained an exact 
equation for the probability of a crossing cluster on a rectangle for 
different aspect ratios, using conformal field theory.

In 1996, Watts extended Cardy's results to obtain an exact equation for
the probability of a cluster crossing both horizontally and vertically
\cite{Watts:1996}. In the same year, after numerical work by Hu and Lin
\cite{HuLin:1996}, Aizenman proved that the probability 
of more than one crossing cluster is finite in the thermodynamic limit
\cite{Aizenman:1997}. This does not contradict the rigorous result that
the number of infinite clusters is either $0$, $1$ or $\infty$ with
probability one \cite{NewmanSchulman:1981a,NewmanSchulman:1981}.  The
asymptote for the probability to obtain $n$ distinct, simultaneously
crossing clusters has been calculated by Cardy \cite{Cardy:1998}, who
also considers spanning on a cylinder.

We define all these probabilities systematically below. In this paper,
we provide accurate numerical data for the many analytical results now
available, many of which are verified for the first time. We also
provide data, particularly for wrapping on the cylinder, for which there
is currently no theoretical explanation.  After briefly discussing the
results for some exotic cluster configurations, we present and discuss results
for crossing, spanning, and wrapping probabilities on rectangles and
cylinders of various aspect ratios for site and bond percolation. 
The results are of such accuracy that we are able to make firm statements as
to the validity of various estimates, conjectures and formulae made in
the past.  An appendix collects together some technical notes on
identifying various cluster types when simulating percolation.

\section{Observables}
In this section we give the various observables that we
measured for site and bond percolation. 
We consider a square lattice in which sites are linked via bonds. To
avoid confusion we stress that boundaries are made of sites only
\cite{LanglandsPichetPouliotStAubin:1992}.
In site percolation sites are occupied with probability $\psite$ 
and all bonds are active. In bond percolation all sites are occupied, 
while bonds are active with probability $\pbond$. Two sites belong to 
the same cluster if they are both occupied and if there is a path 
between them along active bonds and occupied sites. 
A cluster is the set of sites belonging to it.

\subsection{Open boundary conditions}
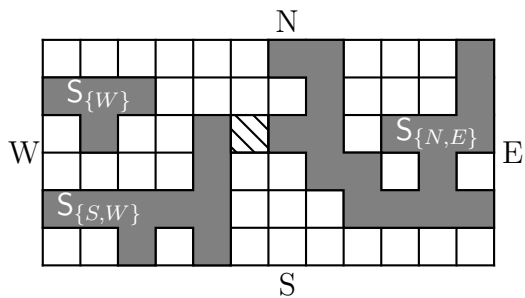
\begin{figure}
\begin{center}
\begin{pspicture}(-0.5,-0.25)(7.5,3.5)

\psgrid[subgriddiv=0,unit=0.5cm,gridlabels=0](0,0)(12,6)

\psset{unit=0.5cm}
\pspolygon[fillstyle=solid,fillcolor=gray]%
(0,1)%
(0,2)%
(4,2)%
(4,4)%
(5,4)%
(5,0)%
(4,0)%
(4,1)%
(3,1)%
(3,0)%
(2,0)%
(2,1)

\rput(1.5,1.5){ {\white $\EventS{12}$} }

\pspolygon[fillstyle=solid,fillcolor=gray]%
(0,4)%
(0,5)%
(3,5)%
(3,4)%
(2,4)%
(2,3)%
(1,3)%
(1,4)

\rput(1.5,4.5){ {\white $\EventS{8}$} }

\pspolygon[fillstyle=solid,fillcolor=gray]%
(6,6)%
(8,6)%
(8,3)%
(9,3)%
(9,2)%
(10,2)%
(10,3)%
(9,3)%
(9,4)%
(11,4)%
(11,6)%
(12,6)%
(12,3)%
(11,3)%
(11,2)%
(12,2)%
(12,1)%
(8,1)%
(8,2)%
(7,2)%
(7,3)%
(6,3)%
(6,4)%
(7,4)%
(7,5)%
(6,5)

\rput(10.5,3.5){ {\white $\EventS{3}$} }

\pspolygon[fillstyle=vlines]%
(5,3)%
(6,3)%
(6,4)%
(5,4)

\rput(12.5,3){E}
\rput(-0.5,3){W}
\rput(6.5,6.5){N}
\rput(6.5,-0.5){S}

\end{pspicture}
\end{center}
\caption{ \label{fig:events} A realisation of a lattice which is
consistent with \fref{fig:prepared_border}.  As indicated, the
clusters are of type $\EventS{3}$, $\EventS{12}$ and $\EventS{8}$. The
hatched square in the centre will be filled to give rise to the border
configuration shown in \fref{fig:wrapping}.}
\end{figure}

Each cluster can be characterised by the borders it touches.  If the
borders of the lattice are labeled $N$, $E$, $S$, $W$, as in
\fref{fig:events}, then each cluster is assigned a subset of these labels,
indicating the borders it touches. There are $15$ different combinations
of cluster labels in which at least one border is touched. In the
following, these combinations are called ``types'',
$\EventYX{S}{\Omega}$.  A cluster of type $\EventYX{S}{\Omega}$ touches
all borders in the set $\Omega$.  For example, the three clusters shown
in \fref{fig:events} are of type $\EventS{3}$, $\EventS{12}$ and
$\EventS{8}$.

We distinguish between clusters touching \emph{only} a set of borders
and \emph{at least} a set of borders. A cluster of type
$\EventYX{S}{\Omega}$ is also of type $\EventYX{U}{\Omega'}$ if $\Omega
\supseteq \Omega'$.  For example, a cluster of type $\EventS{12}$ is
also of type $\EventU{4}$, $\EventU{8}$ and $\EventU{12}$. 
A ``crossing cluster'', i.e. a cluster that connects two
opposite borders, is therefore either of type $\EventU{5}$ or
$\EventU{10}$.\footnote{In the following, we will apply the term
``crossing'' only to systems with open boundaries, and reserve the terms
``spanning'' and ``wrapping'' for cylindrical boundary conditions.}

Normalised histograms $\histo_N(\EventType, n, r)$ have been generated
for all $15$ $\EventS{-1}$ and $\EventU{-1}$ types, where $N$ is the
size of the lattice (number of sites) and $r$ is its aspect ratio
(length over height).  The histogram estimates the probability that a
random realisation contains $n$ clusters of type $\EventType$. It is
worthwhile pointing out that even though $\EventS{-1}$ and $\EventU{-1}$
histograms are correlated, they cannot be derived from each other,
because the number of clusters of a particular $\EventU{-1}$-type must
be determined on a per-realisation basis.

The moments associated with each histogram are defined by
\begin{equation}
        \Moment_N(\EventType, m, r) = \sum_{n=0}^\infty 
        \histo_N(\EventType, n, r) n^m \quad .
\end{equation}
A hat on a quantity, for example $\Momentbar$, indicates its expectation
value, i.e. the value in the limit of an infinitely large ensemble and
system size. 

\subsection{Cylinder}
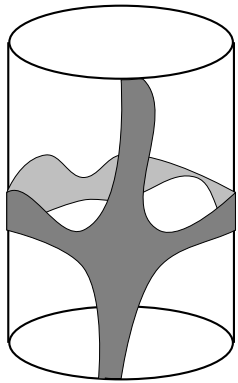
\begin{figure}
\begin{center}
\begin{pspicture}(0,0)(9,5)

\rput(0,-2){
\psellipse(4.5,2)(1.5,0.5)
\psellipse(4.5,6)(1.5,0.5)

\psline(3,2)(3,6)
\psline(6,2)(6,6)

\pscustom[linewidth=0pt]
{
	\psline(2.98,3.5)(2.98,4)
	\pscurve(2.98,4)(3.5,4.5)(4,4.2)(4.5,4.5)(6.02,4)
	\psline(6.02,4)(6.02,3.5)
	\pscurve(6.02,3.5)(5.5,4.2)(4.5,3.9)(4,3.9)(2.98,3.5)
	\fill[fillstyle=solid,fillcolor=lightgray]	
}

\pscustom[linewidth=0pt]
{
	\pscurve(4.2,1.53)(4,3)(2.98,3.5)
	\psline(2.98,3.5)(2.98,4)	
	\pscurve(2.98,4)(4,3.5)(4.5,5.5)
	\psarc(4.5,8){2.5}{270}{275}
	\pscurve(4.8,5.5)(5,3.5)(6.02,4)
	\psline(6.02,4)(6.02,3.5)
	\pscurve(6.02,3.5)(5,3)(4.5,1.52)
	\psarc(4.5,4){2.48}{265}{270}	
	\fill[fillstyle=solid,fillcolor=gray]	
}
}
\end{pspicture}
\end{center}
\caption{\label{fig:cylinder_example} A realisation of percolation on
a cylinder that would contribute to $\histo_N(\WrapSpan,1,r)$ and
$\histo_N(\SpanWrap,1,r)$, but not to $\histo_N(\Wrapping,1,r)$ or
$\histo_N(\Spanning,1,r)$. }
\end{figure}

In addition to the type classification above, we have identified
wrapping and spanning clusters on a cylinder. Our convention for
cylinders is illustrated in \fref{fig:cylinder_example}, which defines
the aspect ratio as circumference over height. A wrapping cluster is
then a cluster which winds around the cylinder, i.e. it provides a path
of length of $2\pi$.  A spanning cluster, meanwhile, connects the bottom
and top of the cylinder.

The resulting histograms are $\histo_N(\Spanning, n, r)$ for the number
distribution of clusters that only span (rather than wrap) the cylinder, 
and $\histo_N(\Wrapping, n, r)$ for the number
distribution of clusters that only wrap (rather than span) the cylinder. 

In addition, $\histo_N(\SpanWrap, 1, r)$ counts the number of realisations
with a single spanning cluster that may also be wrapping. 
Similarly, $\histo_N(\WrapSpan, 1, r)$ counts the number of realisations 
with a single wrapping cluster that may also be spanning.
For an example, see \fref{fig:cylinder_example}. The distinction between
$\Spanning$ and $\SpanWrap$ ($\Wrapping$ and $\WrapSpan$) disappears
for more than one spanning (wrapping) cluster, 
because the existence of two or more simultaneously spanning
(wrapping) clusters prohibits wrapping (spanning) at the same
time. 

\section{Results}
We have generated the histograms mentioned above for site
($\psite=0.59274621$ \cite{NewmanZiff:2000}) and bond percolation
($\pbond=1/2$ \cite{Kesten:1980}), each for three different system
sizes, $N=30000^2, 3000^2, 300^2$. If not mentioned explicitly,
the results presented are for $N=30000^2$. Henceforth, site percolation
is indicated by a superscript $\Ssite$, and bond percolation by
$\Sbond$.  In each of these simulations, $14$ different aspect ratios
were realised while keeping the area constant
\cite{LanglandsPouliotStAubin:1994}.  These aspect ratios are: $30/30$,
$36/25$, $45/20$, $50/18$, $60/15$, $75/12$, $90/10$, $100/9$, $150/6$,
$180/5$, $225/4$, $300/3$, $450/2$ and $900/1$.  Three different boundary 
conditions were applied to each aspect ratio, corresponding to
an open system, a cylinder glued vertically, and a cylinder glued
horizontally. ``Gluing'' is our technical term here referring to the
procedure for applying periodic boundaries. For the definition and
explanation of these technicalities, see
\cite{MoloneyPruessner:2003} and the appendix. The random number
generator used is described in \cite{MatsumotoNishimura:1998b}.
Between $10^6$ and $2 \times 10^6$ independent samples were produced 
for each system size and percolation type, 
using up to $61$ undergraduate computers when idle.

The different aspect ratios are derived from the same set of patches
(s. appendix), by gluing them together to form rectangles in random
permutations and orientations (for details see
\cite{MoloneyPruessner:2003}), and are therefore \emph{not}
statistically independent. However, their correlations are assumed to be
\emph{very} small. When considering the outcome for several aspect
ratios at the same time, one could multiply the error bars by the square
root of the number of aspect ratios considered, as if each aspect ratio
were based only on a subset of the original sample. Because this
procedure assumes maximum correlations between the patches, even though
they are randomly permuted, rotated and mirrored between different
aspect ratios, this is a strong overestimation of the correlations.

We group our results as follows:

\begin{itemize}
\item Corner clusters, types $\EventS{3}$, $\EventS{6}$, $\EventS{12}$
and $\EventS{9}$.  \item ``Three-legged'' clusters, types $\EventS{7}$,
$\EventS{14}$, $\EventS{13}$ and $\EventS{11}$
\item Types $\EventU{5}$ and $\EventU{10}$, the subject of Cardy's
predictions \cite{Cardy:1992,Cardy:1998}, as well as spanning and
wrapping clusters on a cylinder.  
\item Type $\EventS{15}$, in answer to Watts' prediction.
\end{itemize}
Apart from some straightforward arguments, there are no theoretical
predictions for corner and three-legged clusters,
which we will refer to as ``exotic'' types.

\subsection{Exotic types}
\subsubsection{Corner Clusters.}
\begin{figure}[ht]
\begin{center}
\scalebox{0.65}{
\includegraphics{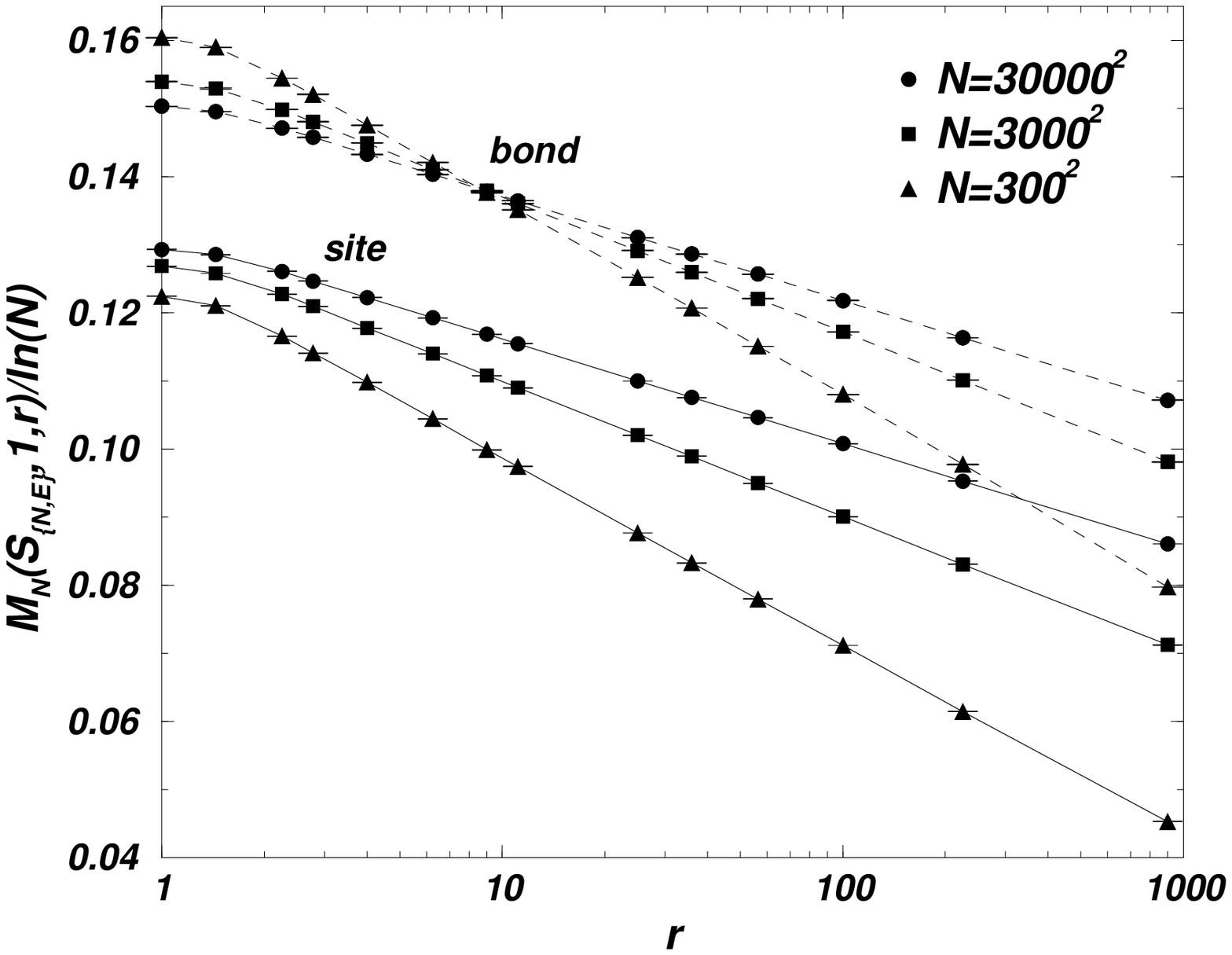}}
\caption{$\Moment_N(\EventS{3}, 1, r)/\ln(N)$ 
for $N=300^2, 3000^2, 30000^2$ and site and bond percolation.
\label{fig:corner_data}
}
\end{center}
\end{figure}

For completeness we include the following results for corner clusters,
i.e. types $\EventS{3}$, $\EventS{6}$, $\EventS{9}$ and $\EventS{12}$.
In a rectangle, all corners are equivalent. Na\"{\i}vely one expects
these clusters to be arranged like onion skins.  Assuming scale
invariance, this suggests a logarithmic dependence of their average
number $\Moment_N(\EventS{-1}, 1, r)$ on the lattice size. This,
however, is not the case, as can be seen in \fref{fig:corner_data},
which should show collapsing lines if $N$ enters only as a factor
$\ln(N)$. It is clear that the graph must level off for $r \to \infty$,
but this region is not yet reached, even for $r=900$ and any $N$. The
lines are remarkably straight, but they seemingly cannot define a
universal exponent.

In the Ising model, the corner magnetisation has been calculated
analytically by Davies and Peschel \cite{DaviesPeschel:1991} using a
corner transfer matrix approach \cite{Baxter:82}. A similar approach
seems to be suitable for corner clusters in percolation.

\subsubsection{Three-legged clusters.} \label{sec:three_leg}
Three-legged clusters touch three borders in a T-like manner. 
There are four different types of three-legged clusters, 
but the symmetry of a rectangle splits them into two pairs, 
while a rotation of the system by $\pi/2$ transforms $r$ into $1/r$ so that
\begin{eqnarray}
        \histobar(\EventS{7}, 1, r)       & = & 
        \histobar(\EventS{13}, 1, r)      = \nonumber \\ 
        \histobar(\EventS{11}, 1, r^{-1}) & = & 
        \histobar(\EventS{14}, 1, r^{-1}) \nonumber \quad .
\end{eqnarray}
The statistics of these clusters turns out to be universal. This is
possibly not very surprising, since they involve a crossing path (see
below). However, what is surprising is the asymptotic value of
$\histo(\EventS{7}, 1, r)$ in the large $r$ limit, when the vertical
crossing probability approaches $1$. At $r=100$ and $N=30000^2$, these 
asymptotic values are $0.5004(4)$ and $0.5002(3)$, 
for site and bond percolation respectively. 
At $r=900$, the values are $0.4999(4)$ and $0.5004(3)$. 
Analysing smaller system sizes reveals that this quantity is
very sensitive to finite-size effects: while even the smallest system
size reaches a value very close to $0.5$ at $r \approx 6$, for small $N$
deviations towards higher probabilities occur at large $r$.  
Evidently, the probability of a three-legged cluster tends to the 
occupation probability of sites on the lattice for $r\to N$, 
where the system effectively becomes a
one-dimensional strip of length $N$ and height $1$.  The region where
the probability remains close to $0.5$ becomes larger as the system
size increases, and for $N=30000^2$ we were unable to detect any
significant deviation from $0.5$ for $r>9$. We therefore conjecture that
\begin{equation}
\lim_{r\to\infty} \histobar(\EventS{7}, 1, r) = 1/2 \quad .
\label{eq:three_leg_assymptote}
\end{equation}
This can be understood as the probability of intersecting a vertically
crossing cluster when vertically cutting a long, narrow percolating
system. However it remains unclear how to derive this limit
analytically. 

\subsection{Results related to Conformal Field Theory}
We now present numerical results that are related to theoretical
predictions based on conformal field theory. This includes measurements
of many formerly unknown quantities, conjectures, and comparisons with
exact results, which all give further support to the conformal
invariance of critical percolation.

\subsubsection{Crossing probability with open boundaries.} 
\label{sec:cross_open} 
Cardy's seminal paper \cite{Cardy:1992} contained a comparison between
his exact result for the crossing probability and the numerical results
obtained by Langland \etal \cite{LanglandsPouliotStAubin:1994}. These
were based on systems with $N=200^2$ sites and $r=1\cdots 7.35$. Later
studies by Shchur and Kosyakov
\cite{ShchurKosyakov:1997,ShchurKosyakov:1998} investigated
$\histo_N(\EventU{5}, n, r)$ at $r=1$ for $n>1$ with very small systems,
$N\le 64^2$. Shchur \cite{Shchur:1999} later extended these results to
systems up to size $N=256\times 3200$, apparently still encountering
finite-size corrections. 
Other studies, such as by Sen
\cite{Sen:1997,Sen:1999} and Hove and Aharony \cite{HoviAharony:1996},
used similar system sizes, while also considering other properties of
spanning clusters.

Using the data presented in this article, it is possible to compare Cardy's
prediction with much greater accuracy, based on systems with $N=30000^2$
sites and $r=1\cdots 900$. However, for very large (and very small)
values of $r$, relevant clusters become either too rare to give any
reasonable estimate for the associated probabilities, or their number
count becomes too broadly distributed. 

The crossing probability is the probability to find at least one
crossing cluster in a particular direction. By symmetry
\begin{equation}
        \histobar(\EventU{5}, n, r) = \histobar(\EventU{10}, n, r^{-1}) 
\end{equation}
which has been used in the data presented below. Consequently,
results for $r>1$ and $r<1$ are not statistically independent,
since they are based on the same realisation (but supposedly different
clusters contribute). Using the short-hand notation 
\begin{equation}
\histo(\EventType, \ge n, r) \equiv 
 \sum_{m=n}^\infty \histo(\EventType, m, r)
\end{equation}
for arbitrary cluster type $\EventType$, Cardy's exact result reads 
\begin{equation}
\histobar(\EventU{5}, \ge 1, r)  =  
  \frac{3\Gamma(\frac{2}{3})}{\Gamma(\frac{1}{3})^2}
  \, \eta^{\frac{1}{3}} \! 
  \phantom{p}_2F_1\left(\frac{1}{3},\frac{2}{3}; \frac{4}{3};\eta \right) 
\label{eq:cardy_at_least} \\
\end{equation}
where
\begin{equation}
        \eta = \left(\frac{1-k}{1+k}\right)^2 \, , 
\quad {\rm and } \quad r^{-1}=\frac{K(1-k^2)}{2K(k^2)} \,
\label{eq:cardy_at_least_appendix} 
\end{equation}
$K(u)$ is the complete elliptic integral of the first kind and
$\phantom{p}_2F_1$ is the hypergeometric function.

\begin{figure}[ht]
\begin{center}
\scalebox{0.65}{ \includegraphics{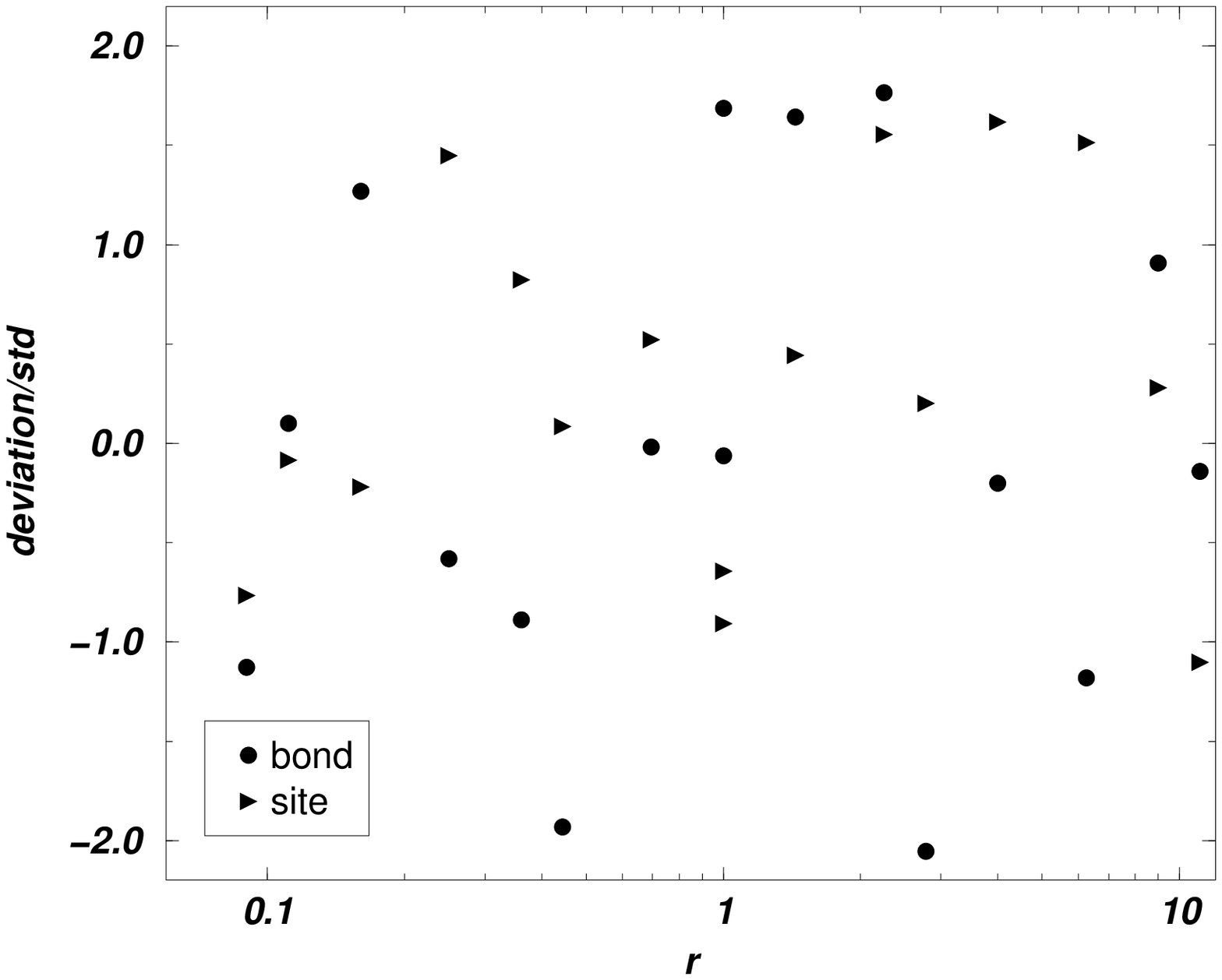}}
\caption{Deviation between analytic (\ref{eq:cardy_at_least}) and
numerical results for the reduced crossing probability
$\histo_N(\EventU{5}, \ge 1, r)$, in units of standard deviations of the
numerical results versus aspect ratio $r$, for bond (circles) and site
(triangles) percolation with $N=30000^2$ sites.
\label{fig:cardy_deviations}}
\end{center}
\end{figure}

\Fref{fig:cardy_deviations} shows the difference between the numerical
result and the exact value from (\ref{eq:cardy_at_least}) in units of
standard deviations. From this plot it is clear that the systematic
deviation for large $r$ observed in
\cite{Cardy:1992,LanglandsPouliotStAubin:1994} was only a finite-size
problem.

The asymptotic number distribution of crossing clusters in percolation
with open boundaries has been derived analytically \cite{Cardy:1998} 
\begin{equation}
        \histobar(\EventU{5}, n, r) \to \exact{\AmpCrossing}(n) 
        \exp\left(-\frac{2}{3} \pi n\left(n - \half\right)/r\right) 
\textrm{ for } n>1 
\label{eq:cardy_flat_exact_n}
\end{equation}
in the limit of $N \to \infty$ and $r \to 0$.  In general, the amplitude
$\AmpCrossing(n)$ is not known exactly, but it is universal and can be
derived from numerics by fitting (\ref{eq:cardy_flat_exact_n}) against
the numerical data in an appropriate region of $r$ values for each $n$
separately. The range of aspect ratios used in the fit is determined by two competing
interests: The fit should include as many points as possible, but exclude
aspect ratios where the asymptotic behaviour has not yet set in.
This determines the largest $r$ in the fit. The smallest $r$ is given by
the value of $r$ for which $n$ simultaneously crossing clusters are
observed at least once.  Fitting ranges are given in all tables
below.

In contrast to spanning on the cylinder, Cardy's asymptotic formula
(\ref{eq:cardy_flat_exact_n}) does not distinguish between a cluster 
crossing in exclusively one direction and a cluster crossing
in possibly more than one direction.  
For consistency in our notation, $\AmpCrossing(n)$ refers to
amplitudes of \emph{exclusively} vertically crossing clusters. 
However, at $n=1$ 
\eref{eq:cardy_flat_exact_n}
applies
to vertically crossing clusters irrespective of other clusters, to which
we assign the amplitude $\exact{\AmpDblCross}(1)$. For $n=1$ Cardy's
prediction (\ref{eq:cardy_flat_exact_n}) therfore reads
\begin{equation}
\histobar(\EventU{5}, 1, r) \to 
  \exact{\AmpDblCross(1)} \exp(-\frac{1}{3} \pi /r) \quad .
\label{eq:cardy_flat_exact_1} 
\end{equation}
There is no prediction for exclusive crossing, 
$\histobar(\EventU{5}, 1, r) - \histobar(\EventU{15}, 1, r)$,
however, it is consistent to fit the latter against
\begin{equation}
  \AmpCrossing(1) \exp(- \alpha(1) /r) \quad .
\label{eq:cardy_flat_exact_simple} 
\end{equation}

The fits throughout the article depend to some degree on the choice of
the fitting interval. \emph{A priori}, it is unknown where the
asymptotic behaviour sets in (in the sense that the deviation from the
asymptote is smaller than the numerical error). It is not possible to
determine whether a deviation of the numerical results from the
analytical value is due to statistical fluctuations or due to a wrongly
chosen fitting interval. The errorbars given can only reflect the
former. However, the error indicated should include the exact result if
it is fitted against the corresponding function in the range given.

\begin{figure}[ht]
\begin{center}
\scalebox{0.65}{
\includegraphics{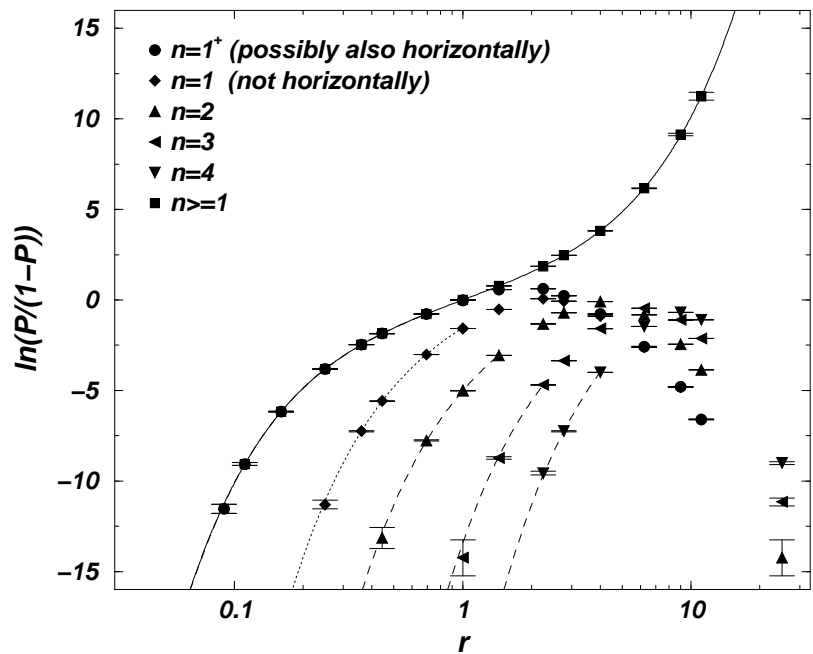}}
\caption{Numerical data ($N=30000^2$, bond percolation, open boundaries) for 
the reduced probability $\ln(\histo/(1-\histo))$ of $n=1,2,3,4$ and $n\ge 1$ vertically 
spanning clusters versus aspect ratio $r$. The data $n\ge 1$ refer to 
$\histo(\EventU{5}, \ge 1, r)$. The long dashed lines are fits according to 
\eref{eq:cardy_flat_exact_n} ((\ref{eq:cardy_flat_exact_1}) and 
(\ref{eq:cardy_flat_exact_simple}) for $n=1$) and 
Tab.~\ref{tab:cardy_flat_amplitudes}, 
while the full line gives the exact result (\ref{eq:cardy_at_least}). 
The dotted line gives the exact result (\ref{eq:cardy_minus_watts}). 
\label{fig:cardy_crossing}
}
\end{center}
\end{figure}

\Fref{fig:cardy_crossing} shows the numerical data for the probability
of finding $n=1,2,3,4$ and $n\ge 1$ vertically crossing clusters in
reduced form, i.e. $\ln(\histo/(1-\histo))$. 
The data are fitted to the asymptotic
formulae (\ref{eq:cardy_flat_exact_n}), (\ref{eq:cardy_flat_exact_1})
and (\ref{eq:cardy_flat_exact_simple}).  As discussed below, the latter
coincides asymptotically with 
\eref{eq:cardy_minus_watts}, 
which
is shown as a dotted line.  For completeness, data for
$\histo(\EventU{5}, \ge 1, r)$ are shown together with the exact result
(\ref{eq:cardy_at_least}).

It is clear that, asymptotically, the relative difference between the
probability of more than one and exactly one crossing cluster,
\begin{equation}
        \frac{\histobar(\EventU{5}, \ge 1, r) - \histobar(\EventU{5}, 1, r)}
        {\histobar(\EventU{5}, \ge 1, r) + \histobar(\EventU{5}, 1, r)}
\label{eq:reldiff_5_5}
\end{equation}
vanishes, and the crossing probability becomes dominated by the probability 
of a single crossing cluster. 
Therefore, the amplitude $\exact{\AmpDblCross}(1)$ in 
(\ref{eq:cardy_flat_exact_1}) is known exactly, namely
\begin{equation}
        \exact{\AmpDblCross}(1) = 2^{4/3} \frac{3 \Gamma(2/3)}{\Gamma(1/3)^2} 
        = 1.4263482556253 \cdots
\label{eq:exact_AmpDblCross}
\end{equation}
using the expansion of \eref{eq:cardy_at_least} provided by Ziff
\cite{Ziff:1995,Ziff:1995_addendum}. 

\Eref{eq:reldiff_5_5} implies that the probability of exactly
one crossing cluster exclusively in one direction, 
$\histobar(\EventU{5}, 1, r) - \histobar(\EventU{15}, 1, r)$,
is the dominating term in the difference 
\begin{eqnarray}
       && \histobar(\EventU{5}, \ge 1, r) - \histobar(\EventU{15}, 1, r)  
 \nonumber \\
       & = & \frac{\eta}{\Gamma(1/3)\Gamma(2/3)} 
       \phantom{p}_3F_2\left(1,1,\frac{4}{3};2,\frac{5}{3};\eta\right) \quad ,
\label{eq:cardy_minus_watts}
\end{eqnarray}
with $\eta$ as in (\ref{eq:cardy_at_least_appendix}).
This relation is derived from (\ref{eq:cardy_at_least}) and 
(\ref{eq:watts_formula}), and for small $r$ can be expanded as
\begin{equation}
        \frac{16}{\Gamma(1/3)\Gamma(2/3)} \exp(-\pi/r) 
        \Big(1 - \frac{8}{5} \exp(-\pi/r) + \frac{4}{3} 
        \exp(-2 \pi/r ) \cdots \Big)
\label{eq:expansion_cardy_minus_watts}
\end{equation}
which is based on Eqn.~(16) in \cite{Ziff:1995} and the definition of
the generalised hypergeometric function. Therefore the two parameters
appearing in (\ref{eq:cardy_flat_exact_simple}) are known exactly:
\begin{equation}
        \exact{\AmpCrossing}(1) =  
        \frac{16}{\Gamma(1/3)\Gamma(2/3)} = 
        4.41063116337433639 \cdots
\label{eq:exact_AmpCrossing1}
\end{equation}
and
\begin{equation}
        \alpha(1) = \pi
\label{eq:exact_alpha1}
\end{equation}
which corresponds to $n=-1$ or $n=3/2$ in (\ref{eq:cardy_flat_exact_n}).

The numerics are in very good agreement with the exact results. 
The estimate for $\AmpDblCross(1)$ shown in 
Tab.~\ref{tab:cardy_flat_amplitudes} agrees perfectly with 
(\ref{eq:exact_AmpDblCross}), while $\AmpCrossing(1)$ has a 
surprisingly large error. However, the result still covers the exact value 
(\ref{eq:exact_AmpCrossing1}). The error is due to a narrow fitting range, 
forced by the late onset of asymptotic behaviour. 
A larger fitting range gives a much smaller error and 
reduced goodness-of-fit \cite{Press:92}. For $\alpha(1)$ it is found
numerically that $\alpha^\Sbond(1)=3.18(6)$ and $\alpha^\Ssite(1)=3.15(5)$,
in perfect agreement with (\ref{eq:exact_alpha1}).

The amplitudes listed in Tab.~\ref{tab:cardy_flat_amplitudes} have also
been used to plot the dashed lines in \fref{fig:cardy_crossing}. They
fit an exponential fairly well (using the $1^\SpWrAdd$ value in
Tab.~\ref{tab:cardy_flat_amplitudes}):
\begin{eqnarray}
  \AmpCrossing^\Sbond(n) & \approx & 0.5422(10) \exp(0.9619(14) n) \\
  \AmpCrossing^\Ssite(n) & \approx & 0.5434(9)  \exp(0.9617(12) n) \quad .
\end{eqnarray}

\begin{table}
\caption{\label{tab:cardy_flat_amplitudes} Crossing with open boundaries. 
Amplitudes are defined as in (\ref{eq:cardy_flat_exact_n}--\ref{eq:cardy_flat_exact_simple}), 
derived from numerical simulations of bond ($\AmpCrossing^\Sbond$) and 
site ($\AmpCrossing^\Ssite$) percolation with $N=30000^2$ sites. 
The $r$ range defines the fitting region (see section \ref{sec:cross_open}). 
The $\SpWrAdd$ marks $\AmpDblCross$, 
i.e. not an exclusively single crossing cluster in one direction.}  
\lineup
\begin{indented}
\item[]\begin{tabular}{c|ll|ll}
\br
$n$          & $r$ range bond        & $\AmpCrossing^\Sbond(n)$ & $r$ range site        & $\AmpCrossing^\Ssite(n)$ \\ 
\mr
 $1^\SpWrAdd$  & $9/100 \cdots 25/36$  &\01.423(2)                & $9/100 \cdots 25/36$  &\01.426(2)  \\
 $1$           & $9/100 \cdots 20/45$  &\04.811(629)              & $9/100 \cdots 25/36$  &\04.560(510)\\
 $2$           & $20/45 \cdots 36/25$  &\03.530(12)               & $20/45 \cdots 36/25$  &\03.553(10) \\
 $3$           & $30/30 \cdots 45/20$  &\09.608(37)               & $30/30 \cdots 45/20$  &\09.599(32) \\
 $4$           & $45/20 \cdots 60/15$  & 27.641(161)              & $45/20 \cdots 60/15$  & 27.658(140) \\
\br
\end{tabular}
\end{indented}
\end{table}

\subsubsection{Spanning probability on a cylinder.}
Cardy has also given asymptotes for spanning events on a cylinder
\cite{Cardy:1998}. These clusters have been investigated numerically
several times \cite{HoviAharony:1996,ShchurKosyakov:1997,Shchur:1999}.
However, this work uses system sizes three orders of magnitude larger 
than in former studies. The probability of obtaining $n$ distinct
spanning clusters on a cylinder is expected to behave in the limit of
small $r$ and large $L$ like
\begin{equation}
\histobar(\Spanning, n, r) \to 
  \AmpSpanning(n) \exp(-\frac{2}{3} \pi (n^2 - \quarter)/r) \quad .
\label{eq:cardy_cylinder_exact_n}
\end{equation}
The existence of a wrapping cluster prevents more than one spanning cluster, 
see \fref{fig:cylinder_example}. For $n=1$ one can distinguish between 
exclusively spanning clusters and spanning clusters that may also wrap. 
Allowing for wrapping clusters Cardy predicts the asymptote
\begin{equation}
\histobar(\SpanWrap, 1, r) \to 
  \AmpSpanWrap(1) \exp(-\frac{5}{24} \pi /r) \quad .
\label{eq:cardy_cylinder_possibly_wrap}
\end{equation}

The numerical results are in full agreement with 
\eref{eq:cardy_cylinder_exact_n} and \eref{eq:cardy_cylinder_possibly_wrap}.
The corresponding amplitudes
are shown in Tab.~\ref{tab:cardy_cylinder_amplitudes}.  The fact that
the amplitude in (\ref{eq:cardy_cylinder_possibly_wrap}),
$\AmpSpanWrap(1)$, is slightly smaller than the corresponding amplitude
in (\ref{eq:cardy_cylinder_exact_n}), $\AmpSpanning(n)$, does not
contradict $\histobar(\Spanning, 1, r) < \histobar(\SpanWrap, 1, r)$,
since the exponentials differ by a factor $\exp(\pi (7/24)/r)$ in favour
of spanning without restrictions on wrapping. It would require an
$r>2.67$ to suppress this factor enough to equalise both probabilities.
However, at such large values of $r$, Eqn.'s
(\ref{eq:cardy_cylinder_exact_n}) and
(\ref{eq:cardy_cylinder_possibly_wrap}) are not valid any longer.

\begin{table}
\caption{\label{tab:cardy_cylinder_amplitudes} Spanning on a
cylinder. Amplitudes are defined as in
(\ref{eq:cardy_cylinder_exact_n}), where $\SpWrAdd$ marks the value of
$\AmpSpanWrap(1)$ (see (\ref{eq:cardy_cylinder_possibly_wrap})).}
\lineup
\begin{indented}
\item[]\begin{tabular}{c|ll|ll}
\br
$n$          & $r$ range bond       & $\AmpSpanning^\Sbond(n)$ & $r$ range site       & $\AmpSpanning^\Ssite(n)$ \\
\mr
$1^\SpWrAdd$   & $10/90 \cdots 30/30$ &\01.2217(4)               & $10/90 \cdots 30/30$ &\01.2222(4) \\
$1$            & $9/100 \cdots 30/30$ &\01.7198(11)              & $9/100 \cdots 30/30$ &\01.7225(10)\\
$2$            & $25/36 \cdots 36/25$ &\05.1829(256)             & $25/36 \cdots 36/25$ &\05.2105(218)\\
$3$            & $36/25 \cdots 50/18$ & 15.1212(764)             & $30/30 \cdots 50/18$ & 15.0227(649)\\
$4$            & $45/20 \cdots 60/15$ & 45.0059(3280)            & $45/20 \cdots 60/15$ & 44.5445(2780)\\
\br
\end{tabular}
\end{indented}
\end{table}

The amplitudes fit an exponential extremely well 
(where for $n=1$ wrapping was not allowed, i.e. the result $1^\SpWrAdd$ in 
Tab.~\ref{tab:cardy_cylinder_amplitudes} was ignored):
\begin{eqnarray}
       \AmpSpanning^\Sbond(n) & \approx & 0.5788(11) \exp(1.0892(17) n) \\
       \AmpSpanning^\Ssite(n) & \approx & 0.5816(10) \exp(1.0860(14) n) \quad .
\end{eqnarray}

The resulting plot of the reduced probabilities looks very similar to
\fref{fig:cardy_crossing}, so we omit it here.

\subsubsection{Wrapping on a cylinder.}
\begin{figure}[ht]
\begin{center}
\scalebox{0.65}{
\includegraphics{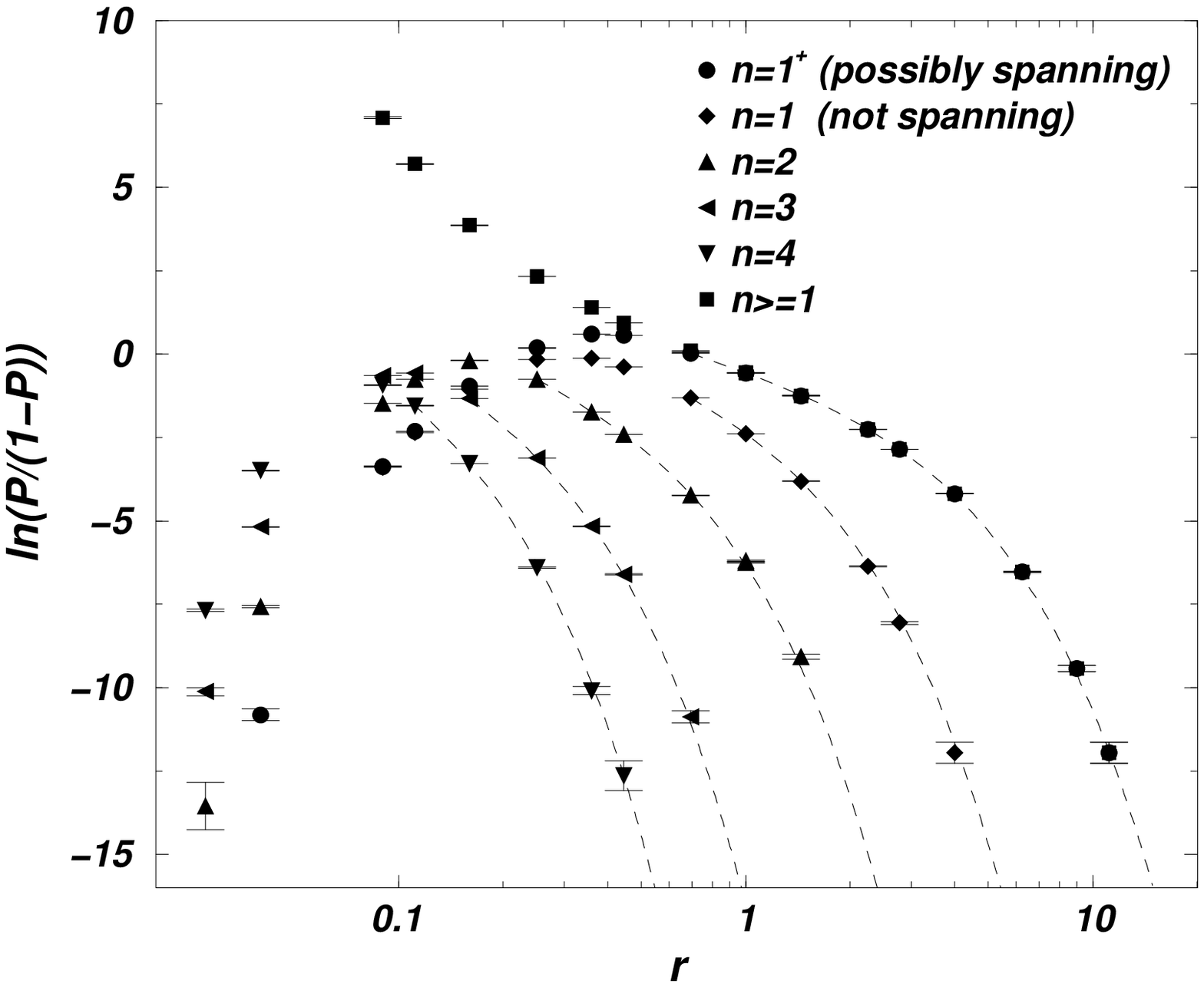}}
\caption{Numerical data ($N=30000^2$, bond percolation) for the reduced probability 
$\ln(\histo/(1-\histo))$ of $n=1,2,3,4$ and $n\ge 1$ wrapping clusters on a cylinder
versus aspect ratio $r$. The data $n\ge 1$ refer to $\histo(\Wrapping, \ge 1, r)$ 
and look qualitatively similar to those shown in \fref{fig:cardy_crossing}.
The long dashed lines are fits  according to \eref{eq:wrapping_fit} and Tab.~\ref{tab:wrapping}.
\label{fig:wrapping_data}
}
\end{center}
\end{figure}

Aizenman's original statement \cite{Aizenman:1997} regarding the number
of crossing clusters can also be applied to the number of wrapping
clusters on a cylinder. In the limit of large $r$ 
\begin{equation}
        \ln(\histobar(\Wrapping, n, r)) \in \OC(r n^2)
\end{equation}
according to a hand-waving scaling argument by Cardy \cite{Cardy:1998}. 
There is no better estimate known, so by fitting each set of histograms to 
\begin{equation}
        \histobar(\Wrapping, n, r) = \AmpWrapping(n) \exp(\alpha_\Wrapping(n) r)
\label{eq:wrapping_fit}
\end{equation}
one can determine the $n$ dependence of $\alpha_\Wrapping(n)$ and the
amplitude $\AmpWrapping(n)$. Tab.~\ref{tab:wrapping} shows the
corresponding results. It turns out that $\alpha_\Wrapping(n)$ fits very
well a second order polynomial with coefficients
\begin{eqnarray}
 \alpha^\Sbond_\Wrapping(n) & \approx & -3.150(11) n^2 + 5.51(4) n - 5.41(3) \\
 \alpha^\Ssite_\Wrapping(n) & \approx & -3.176(10) n^2 + 5.61(4) n - 5.48(3)
 \quad .
\end{eqnarray}
The amplitude $\AmpWrapping(n)$ again fits an exponential, but
only if the very first value, $\AmpWrapping(1)$, is neglected. In this
case we find
\begin{eqnarray}
        \AmpWrapping^\Sbond(n) & \approx & 0.483(5) \exp(0.662(4) n) \\
        \AmpWrapping^\Ssite(n) & \approx & 0.473(4) \exp(0.670(3) n) \quad .
\end{eqnarray}
We stress again that the ambiguity in the choice of the fitting ranges 
introduces an error which is not reflected in the numerical error given. 
We find it therefore justified to conjecture that 
\begin{equation}
        \alpha_\Wrapping(n) = - \pi n^2 + \frac{7}{4} \pi (n-1)
\label{eq:wrapping_conjecture}
\end{equation}
which is somewhat surprising, since we na\"{\i}vely expected 
similar arguments \cite{Cardy:1998}
to those for (\ref{eq:cardy_cylinder_exact_n}) should apply, 
giving rise to a leading term $- \frac{2}{3} \pi n^2$. 
The numerical results are shown together with the proposed analytical 
behaviour according to (\ref{eq:wrapping_fit}) using the data in 
Tab.~\ref{tab:wrapping} in \fref{fig:wrapping_data}.

\begin{table}
\caption{\label{tab:wrapping} Wrapping on a cylinder. Results for the
fits of $\histo(\Wrapping, n, r)$ to \eref{eq:wrapping_fit}.  The
$r$ range defines the fitting region, where the right-hand value is for
the largest aspect ratio in which $n$ wrapping clusters occured at least
once.  The $\SpWrAdd$ marks the value of $\AmpWrapSpan(1)$ that allows
a spanning cluster. } \lineup
\begin{tabular}{l|lll|lll}
\br
$n$& $r$ range (bond)       & $\AmpWrapping^\Sbond(n)$ & $\alpha^\Sbond_\Wrapping(n)$ & $r$ range (site)       & $\AmpWrapping^\Ssite(n)$ & $\alpha^\Ssite_\Wrapping(n)$ \\
\mr
$1^\SpWrAdd$ & $25/36 \cdots 100/9$ & 1.0654(12)               &\0-1.0755(10)                 & $25/36 \cdots 100/9$ & 1.0634(10)               &\0-1.0739(9)  \\
$1$          & $25/36 \cdots 60/15$ & 1.7734(82)               &\0-3.0536(52)                 & $25/36 \cdots 60/15$ & 1.7712(70)               &\0-3.0522(44) \\
$2$          & $15/60 \cdots 36/25$ & 1.8134(55)               &\0-6.9320(92)                 & $15/60 \cdots 45/20$ & 1.7995(46)               &\0-6.9118(78) \\
$3$          & $12/75 \cdots 25/36$ & 3.6128(216)              & -17.8021(326)                & $12/75 \cdots 30/30$ & 3.6600(186)              & -17.8764(278) \\
$4$          & $10/90 \cdots 20/45$ & 6.6734(609)              & -32.6919(752)                & $10/90 \cdots 20/45$ & 6.6533(520)              & -32.7071(645) \\
\br
\end{tabular}
\end{table}

\subsubsection{Spanning simultaneously in both directions.}
Watts \cite{Watts:1996} has exactly calculated the probability of a cluster 
that crosses both directions simultaneously in a system with open boundaries, 
$\histobar(\EventU{15}, 1, r)$, given by 
\begin{eqnarray}
\histobar(\EventU{15}, 1, r) = && \label{eq:watts_formula} \\
\frac{3\Gamma(\frac{2}{3})}{\Gamma(\frac{1}{3})^2} \, \eta^{\frac{1}{3}} \! 
\phantom{p}_2F_1\left(\frac{1}{3},\frac{2}{3}; \frac{4}{3};\eta \right) 
-
\frac{\eta}{\Gamma(\frac{1}{3}) \Gamma(\frac{2}{3})} 
\phantom{p}_3F_2\left(1,1,\frac{4}{3};2,\frac{5}{3};\eta\right) \quad , 
\nonumber
\end{eqnarray}
with $\eta$ as defined in (\ref{eq:cardy_at_least_appendix}) and 
$\phantom{p}_3F_2$ being the generalised hypergeometric function.
The first term on the right-hand side of (\ref{eq:watts_formula}) 
is identical to $\histobar(\EventU{5}, \ge 1, r)$. According to the expansion
provided by Ziff \cite{Ziff:1995} and consistent with 
(\ref{eq:cardy_flat_exact_1}), for small $r$ this term is proportional to 
$\exp(-\frac{1}{3} \pi /r)$, while, according to
(\ref{eq:expansion_cardy_minus_watts}), the second term decays even
faster, namely $\exp(-\pi /r)$. This is of course what one expects, because
the probability of spanning in both directions is, away from $r=1$,
dominated by the probability of spanning in the longer direction.

\begin{figure}[ht]
\begin{center}
\scalebox{0.65}{ \includegraphics{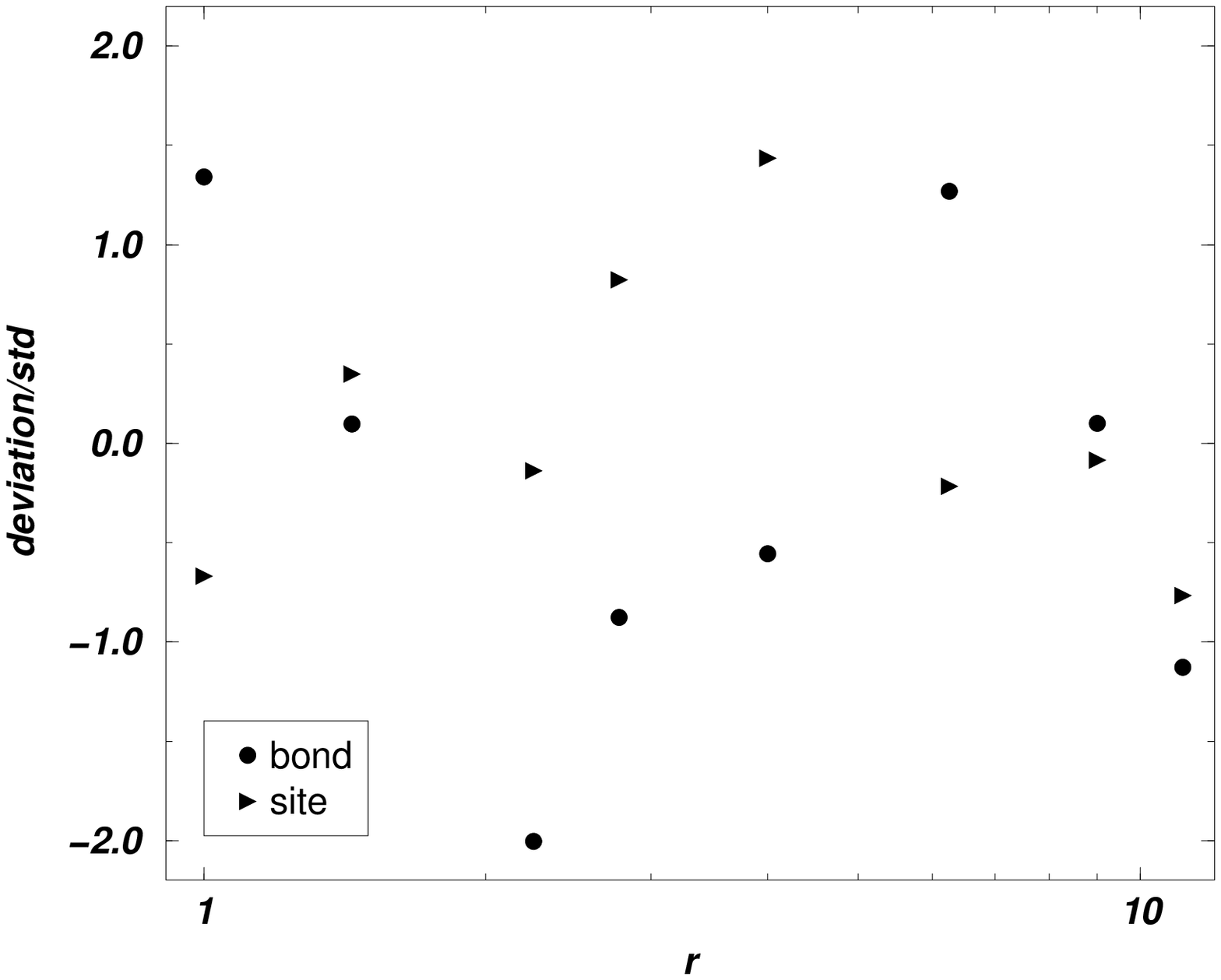}}
\caption{Deviation between the analytic result (\ref{eq:watts_formula})
and the numerical results for the reduced crossing probability
$\histo_N(\EventU{15}, 1, r)$, in units of standard deviations of the
numerical results versus aspect ratio $r$, for bond (circles) and site
(triangles) percolation with $N=30000^2$ sites.
\label{fig:watts_deviations}}
\end{center}
\end{figure}

\Fref{fig:watts_deviations} shows the difference between our numerical
results and Watts' prediction. As in \fref{fig:cardy_deviations}, the
plot suggests that the deviations observed in \cite{Watts:1996} are only
finite-size corrections.

\subsection{Finite-size corrections}
Since the system sizes investigated are huge compared to former studies
\cite{LanglandsPouliotStAubin:1994,ShchurKosyakov:1997,Shchur:1999}, one
might be inclined to completely ignore finite-size corrections.  
In order to estimate their strength, Tab.~\ref{tab:compare} lists numerical
results of various quantities for different system sizes and compares
them with estimates for the value of these quantities in the
thermodynamic limit found in the literature or their exact results. 
For large systems, our numerical results agree very well. 
For smaller systems, there may be some mild corrections. 
From the three-legged cluster data (sec. \ref{sec:three_leg}) we expect that
finite-size corrections may become visible if one side of the
rectangle has a length $\le 300$, which is not the case for any aspect
ratio we have simulated in a system with $N=30000^2$ sites.

The results also indicate that the estimate of $p_c^\Ssite$ by Newman
and Ziff \cite{NewmanZiff:2000} is valid within numerical error.  This
is corroborated by \fref{fig:cardy_deviations} and
\fref{fig:watts_deviations}, where site percolation does not seem to
show stronger deviations than bond, for which $p_c^\Sbond$ is known
exactly.

\begin{table}
\caption{\label{tab:compare} Comparison of our numerical values and
estimates found in the literature, for different system sizes.
``Estimates'' are values for those quantities cited in the literature
in the thermodynamic limit.}  \lineup
\begin{indented}
\item[]\begin{tabular}{l|ll|l}
\br
Quantity                                     & site                    & bond     & Exact value/Estimate \\
\mr
$\histo_{N=3000^2}(\EventU{5},   \ge 1, 1)$  & $0.4995(4)$               & $0.5012(4)$             &     \\
$\histo_{N=30000^2}(\EventU{5},  \ge 1, 1)$  & $0.4998(3)$               & $0.4999(4)$             &  
    \raisebox{1.5ex}[-1.5ex]{$1/2$ \textrm{ exact} \cite{LanglandsPichetPouliotStAubin:1992}} \\
\mr
$\histo_{N=3000^2}(\EventU{5},   \ge 2, 1)$  & $6.58(6) \times 10^{-3}$   & $6.69(7) \times 10^{-3}$ &      \\
$\histo_{N=30000^2}(\EventU{5},  \ge 2, 1)$  & $6.71(6) \times 10^{-3}$   & $6.66(6) \times 10^{-3}$ &  
    \raisebox{1.5ex}[-1.5ex]{$6.58(3) \times 10^{-3}$ \cite{ShchurKosyakov:1997}} \\
\mr
$\histo_{N=3000^2}(\EventU{15},   1, 1)$     & $0.3221(4)$               & $0.3230(4)$             &      \\
$\histo_{N=30000^2}(\EventU{15},  1, 1)$     & $0.3219(3)$               & $0.3226(4)$             &  
    \raisebox{1.5ex}[-1.5ex]{$0.322120455\cdots$ \textrm{ exact}\cite{Watts:1996}} \\
\mr
$\histo_{N=3000^2}(\Spanning,    \ge 1, 1)$  & $0.6360(4)$               & $0.6368(4)$             &      \\
$\histo_{N=30000^2}(\Spanning,   \ge 1, 1)$  & $0.6361(3)$               & $0.6364(4)$             &  
    \raisebox{1.5ex}[-1.5ex]{$0.63665(8)$ \cite{HoviAharony:1996}}\\
\br
\end{tabular}
\end{indented}
\end{table}

\section{Conclusion and Discussion}
Our numerical results represent probably the most comprehensive and most
up-to-date study of crossing, spanning and wrapping probabilities. We
have presented results for ``exotic'' cluster types, for which an
analytical description is still lacking.  With regards to the
predictions of conformal field theory, our numerical data affords a
comparison between numerical and analytical results, and we give further
support to conformal invariance in percolation and other critical
phenomena \cite{Cardy:1987}.  We believe that deviations from the
predicted behaviour observed in the literature are most likely due to
finite-size effects.

We have also calculated the amplitudes listed in
Tab.~\ref{tab:cardy_flat_amplitudes},
Tab.~\ref{tab:cardy_cylinder_amplitudes} and Tab.~\ref{tab:wrapping},
which might be of help to theorists. For three-legged clusters, we have
conjectured the asymptote (\ref{eq:three_leg_assymptote}), for wrapping
probabilities the form (\ref{eq:wrapping_conjecture}).

\ack{The authors wish to thank Andy Thomas for his fantastic technical
support.  Without his help and dedication, this project would not have
been possible.  The authors also thank Dan Moore, Brendan Maguire and
Phil Mayers for their continuous support, as well as Kim Christensen
for his helpful comments.  NRM is very grateful to the Beit Fellowship,
and to the Zamkow family.  GP gratefully acknowledges the support of the
EPSRC.}

\section*{Appendix}
Here we list a few technicalities on how to identify cluster types, etc. 
The identification methods are applicable in any simulation of percolation 
where the boundary can be represented in the Hoshen-Kopelman
\cite{HoshenKopelman:1976} form.

\subsection*{Simulation Method}
Our method performs asynchronously parallelised percolation on
distributed machines \cite{MoloneyPruessner:2003}. 
In principle, the method relaxes all the standard constraints in numerical 
simulations of percolation, such as CPU power, memory, and network capacity.  
It is especially suited for calculating cluster size distributions and 
finite-size corrections, crossing probabilities, and, by applying the
corresponding boundary conditions, distributions of wrapping and
spanning clusters on different topologies, e.g. cylinder, torus, or the
M\"{o}bius strip.

\begin{figure}
\begin{center}
\begin{pspicture}(-0.5,-0.25)(7.5,3.5)

\psframe[fillstyle=solid,fillcolor=lightgray,linecolor=lightgray](0,0)(6,3)
\psgrid[subgriddiv=0,unit=0.5cm,gridlabels=0](0,0)(12,6)
\psframe[fillstyle=solid,fillcolor=white,linecolor=white]    (0.52,0.52)(5.48,2.48)

\psset{unit=0.5cm}
\rput(0.5,0.5){
\rput(0,5){{\small 0}}
\rput(1,5){{\small 0}}
\rput(2,5){{\small 0}}
\rput(3,5){{\small 0}}
\rput(4,5){{\small 0}}
\rput(5,5){{\small 0}}
\rput(6,5){{\small -17}}
\rput(7,5){{\small 7}}
\rput(8,5){{\small 0}}
\rput(9,5){{\small 0}}
\rput(10,5){{\small 0}}
\rput(11,5){{\small 7}}

\rput(11,4){{\small 12}}
\rput(11,3){{\small 13}}
\rput(11,2){{\small 0}}
\rput(11,1){{\small {\white 13}}}
\rput(11,0){{\small 0}}

\rput(0,0){{\small 0}}
\rput(1,0){{\small 0}}
\rput(2,0){{\small 24}}
\rput(3,0){{\small 0}}
\rput(4,0){{\small -9}}
\rput(5,0){{\small 0}}
\rput(6,0){{\small 0}}
\rput(7,0){{\small 0}}
\rput(8,0){{\small 0}}
\rput(9,0){{\small 13}}
\rput(10,0){{\small 0}}
\rput(11,0){{\small 0}}

\rput(0,4){{\small -4}}
\rput(0,3){{\small 0}}
\rput(0,2){{\small 0}}
\rput(0,1){{\small {\white 24}}}
\rput(0,0){{\small 0}}
}

\rput(0.5,0){
\psbezier{->}(7,5)(7,4.5)(6,4.5)(6,5)
\psbezier{->}(11,6)(9,7)(6,7.7)(6,6)

\psbezier{->}(2,1)(2,2)(3,2)(4,1)
\psbezier{->}(0.5,1.5)(2,2.5)(3,2.5)(4,1)
\pscurve[linestyle=dashed]{->}(4,0)(3.5,-0.2)(-0.4,-0.4)(-0.7,6)(5.3,6.3)(5.7,6)
}

\rput(0,0.5){
\psbezier{->}(12,4)(13.2,5.5)(12,6.5)(11.5,5.5)

\psbezier{->}(11,3)(10.5,3)(10.5,4)(11,4)
\psbezier{->}(12,1)(13.5,3)(13,4)(12,4)
}

\end{pspicture}
\end{center}
\caption{ \label{fig:prepared_border} An example of a labeled border. 
A negative entry indicates a cluster size, a positive
entry is a pointer to another site, and a zero entry indicates that the
site is not occupied. The configuration shown is a border
representation of a full lattice as in \fref{fig:events}. 
The dashed pointer applies for periodic boundary conditions along the
vertical borders, joining the two highlighted sites.}
\end{figure}
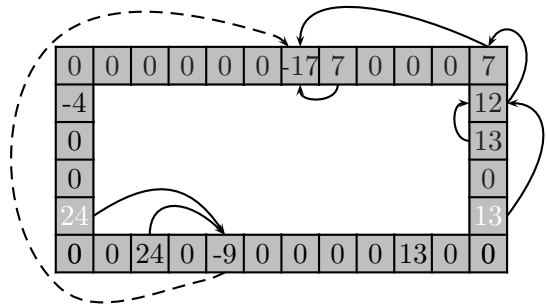

The method is based on a master/slave parallelisation, where slaves send
``patches'' (specially prepared borders representing the lattice) to a
master node, which ``glues'' these patches together. After a path
compression, which is essentially a form of ``Nakanishi label
recycling'' \cite{NakanishiStanley:1980,BinderStauffer:1987} where bulk
sites are considered inactive, the result is a single border as shown in
\fref{fig:prepared_border}. The term ``gluing'' will be used to indicate
that a configuration is updated to account for a link introduced between
two sites or two boundaries. The Hoshen-Kopelman (HK) algorithm
\cite{HoshenKopelman:1976} provides the data representation, which is
the key of this method. Extremely large system sizes can be simulated.
 
An example of a border representation is shown in
\fref{fig:prepared_border} (in the following all examples are based on
site percolation). Each site on the border is indexed in a clockwise
manner starting with $1$ in the upper left corner. Each site also
contains a label. If the label is $0$ the site is not occupied. If it is
positive it is interpreted as a ``pointer'' to the index of another site
in the same cluster.  If it is negative, it is called a ``root'', and the
magnitude of the negative number indicates the size of the cluster the
site belongs to. The HK algorithm ensures that all sites of a cluster
form a single tree, with a root that carries a negative label.  In this
way it is possible to identify the cluster each site belongs to by its
corresponding root.  All information about a cluster is stored at the
root site.

It is technically simple to identify the types of all clusters touching the 
border: The algorithm scans along a border like the one shown in 
\fref{fig:prepared_border} and assigns a flag to the root of each (occupied) 
site it is visiting, according to the location of the site visited. 
The flags gathered at each root then identify the borders that the represented 
clusters touch. The $\histo_N(\EventType, n, r)$ are based on the
statistics of these flags.

\subsection*{Identifying spanning and wrapping} 
Starting from a configuration with open boundaries (for example
\fref{fig:prepared_border}), spanning clusters on a cylinder can be
detected easily, by ``gluing'' the appropriate borders: First the roots
are identified for each pair of sites, which become nearest neighbours due
to the new boundary conditions.  Preferably, the root of the smaller
cluster is then redirected to the other, i.e. its label is overwritten
by the index of the other root, which, in turn, inherits all properties
of the overwritten root, such as cluster size or border flags.  The new
pointer resulting from a vertical gluing is shown as a dashed line in
\fref{fig:prepared_border}; clusters of type $\EventU{5}$ are then
spanning on the cylinder.

It is significantly more complicated to detect wrapping clusters, because
these clusters cannot be defined by properties of individual sites alone.

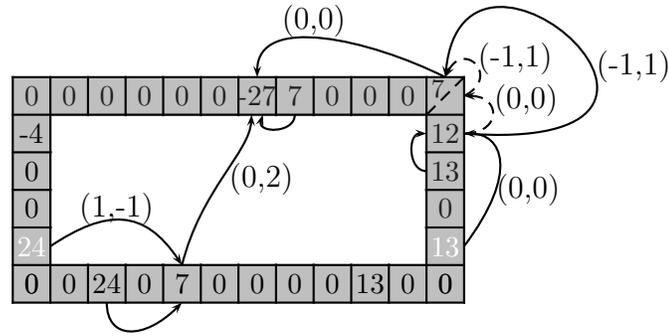
\begin{figure}
\begin{center}
\begin{pspicture}(-0.5,-0.25)(7.5,3.8)

\psframe[fillstyle=solid,fillcolor=lightgray,linecolor=lightgray](0,0)(6,3)
\psgrid[subgriddiv=0,unit=0.5cm,gridlabels=0](0,0)(12,6)
\psframe[fillstyle=solid,fillcolor=white,linecolor=white]    (0.52,0.52)(5.48,2.48)

\psset{unit=0.5cm}
\rput(0.5,0.5){
\rput(0,5){{\small 0}}
\rput(1,5){{\small 0}}
\rput(2,5){{\small 0}}
\rput(3,5){{\small 0}}
\rput(4,5){{\small 0}}
\rput(5,5){{\small 0}}
\rput(6,5){{\small -27}}
\rput(7,5){{\small 7}}
\rput(8,5){{\small 0}}
\rput(9,5){{\small 0}}
\rput(10,5){{\small 0}}
\rput(10.8,5.2){{\footnotesize 7}}

\rput(11,4){{\small 12}}
\rput(11,3){{\small 13}}
\rput(11,2){{\small 0}}
\rput(11,1){{\small {\white 13}}}
\rput(11,0){{\small 0}}

\rput(0,0){{\small 0}}
\rput(1,0){{\small 0}}
\rput(2,0){{\small 24}}
\rput(3,0){{\small 0}}
\rput(4,0){{\small 7}}
\rput(5,0){{\small 0}}
\rput(6,0){{\small 0}}
\rput(7,0){{\small 0}}
\rput(8,0){{\small 0}}
\rput(9,0){{\small 13}}
\rput(10,0){{\small 0}}
\rput(11,0){{\small 0}}

\rput(0,4){{\small -4}}
\rput(0,3){{\small 0}}
\rput(0,2){{\small 0}}
\rput(0,1){{\small {\white 24}}}
\rput(0,0){{\small 0}}

}

\rput(0.5,0){
\psbezier{->}(7,5)(7,4.5)(6,4.5)(6.15,5)

\psbezier{->}(11,6)(9,7)(6,7.7)(6,6)
\rput(7.5,7.5){(0,0)}

\psbezier{->}(2,0)(2,-1)(3,-1)(4,0)
\psbezier{->}(0.5,1.5)(2,2.5)(3,2.5)(4,1)
\rput(2.25,2.5){(1,-1)}

\psbezier{->}(4,1)(4.5,3)(5.8,4)(5.85,5)
\rput(6.1,3.3){(0,2)}

}

\rput(0,0.5){
\psbezier[linestyle=dashed]{->}(12,4)(13,4)(13,5)(12,5)
\rput(13.65,4.9){(0,0)}

\psbezier[linestyle=dashed]{->}(12,5)(13,5.5)(12,6.5)(11.5,5.5)
\psline[linestyle=dashed](12,5.5)(11,4.5)
\rput(13.35,6.0){(-1,1)}

\psbezier{->}(12,4)(20,3.4)(12,10.5)(11.5,5.5)
\rput(16.5,5.8){(-1,1)}

\psbezier{->}(11,3)(10.5,3)(10.5,4)(11,4)
\psbezier{->}(12,1)(13.5,3)(13,4)(12,4)
\rput(13.7,2.5){(0,0)}
}

\end{pspicture}
\end{center}
\caption{ \label{fig:wrapping} The HK representation of the border of the 
configuration shown in \fref{fig:events} with the hatched square ``activated'' 
to allow for a wrapping cluster. The tuples indicate the distance of a site to 
the site it is pointing to. These numbers are only given for the two paths 
which are relevant for the emergence of a wrapping cluster when the system is 
``glued'' along the $W$ and $E$ borders. The dashed line in the top right-hand 
corner indicates the special treatment of a corner, while the dashed arrows 
indicate auxiliary pointers. The highlighted sites give rise to a wrapping 
cluster.}
\end{figure}

The simplest solution, which is applicable to any kind of topology, such
as a cylinder, torus or M\"{o}bius strip, is to assign to each site an
additional set of numbers, which indicates its distance to the site it
is pointing to. This distance can be measured in units of lattice
spacings or, even simpler, in multiples of $\pi$: each type of border is
mapped to a set of integers like
\begin{equation}
        N \to (0,1) \quad S \to (0,-1) \quad E \to (1,0) \quad W \to (-1,0)
\end{equation} 
and for each site an additional flag indicates its location. The
distance between two sites is then the difference between the tuples
associated with their location flag. For example, the distance between
$W$ and $S$ is $(1,-1)$. In the open system, each site is assigned a
tupel indicating the distance to the site it is pointing to.  The
periodic boundary conditions are then applied by gluing along the
vertical borders. Two clusters merge in the way described above, with
the additional assignment of a distance vector to the redirected root
site. This distance vector indicates the distance from the redirected
root site to the new root site, given by the difference between the
distances of the glued sites and their roots. In that way, for any site
in a cluster the distance to the root is given by the sum over all
distance vectors along the tree to the root.

If the gluing procedure now comes across two sites on either border,
which belong to the same cluster\footnote{This in itself does not indicate
that a cluster is wrapping, as a previous encounter of the two clusters
may have merged them.}, the pathlength to the root is calculated for
each of them. If their difference is non-zero, a wrapping cluster has
been found, the number indicating the winding number as a multiple of
$\pi$. In the example in \fref{fig:wrapping}, the left hand path has
length $(1,-1)+(0,2)=(1,1)$, the right hand path $(-1,1)$, differing by
the expected length $(0,2)$. It is a topological fact that higher
winding numbers cannot appear on a cylinder.

Corners require special attention, because they belong to two different
borders at the same time. In the example shown in \fref{fig:wrapping},
all paths shown on the right-hand side have length $(0,0)$, apart from
the corner site, which is thought as carrying an ``internal pointer'' to
itself, connecting the $E$ border to the $N$ border.  The internal
pointer can be understood as follows: if a site points to a corner, it
has a connection to two borders at once. However, the distance vector of
the pointing site can only indicate the distance to one border.  A
convention is required to lift this degeneracy, for example that all
pointers pointing towards and away from a corner site do so with respect
to the $N$ or $S$ boundary, never with respect to the $E$ or $W$
boundary. However, it might happen that another corner is glued to the
corner site, or a neighbouring site is connected to it \emph{as a $W$ or
$E$ site}. In this case, auxiliary pointers are introduced: one pointer
from the connected site to the corner site's $W$ or $E$ part and another
pointer internally connecting the corner site's $W$ or $E$ part to the
$N$ or $S$ part. In \fref{fig:wrapping} these auxiliary pointers and the
sub-partitioning of a corner site are shown with dashed lines.

\section*{References}
\bibliography{articles,books}
\end{document}